\documentclass[a4paper,11pt]{article}

\usepackage{amsmath}
\usepackage{amssymb}
\usepackage{fancyhdr}
\usepackage{geometry}
\usepackage{graphicx}
\usepackage{natbib}
\usepackage{times}

\title{\bf The importance of selection rate \\
in the evolution of cooperation}

\author{Carlos P. Roca$^{1,2}$,
Jos\'e A. Cuesta$^1$ and Angel S\'anchez$^{1,3}$
\thanks{Corresponding author. Tel: +34 916249411.
Fax: +34 916249129. E-mail: anxo@math.uc3m.es}\\[3ex]
\em $^1$Grupo Interdisciplinar de Sistemas Complejos (GISC)\\
\em Departamento de Matem\'aticas,
Universidad Carlos III de Madrid\\
\em 28911 Legan\'es, Madrid, Spain\\
\em $^2$Centro Nacional de Biotecnolog\'\i a - CSIC,
28049 Madrid, Spain\\
\em $^3$Instituto de Biocomputaci\'on y F\'\i sica de Sistemas
Complejos (BIFI)\\
\em Universidad de Zaragoza, 50009 Zaragoza, Spain\\[3ex]}

\begin{document}

\maketitle

\newpage

\begin{abstract}
How cooperation emerges in human societies is still a puzzle.
Evolutionary game theory has been the standard framework to address
this issue. In most models, every individual plays with all others,
and then reproduce and die according to what they earn. This amounts
to assuming that selection takes place at a slow pace with respect to
the interaction time scale. We show that, quite generally, if
selection speeds up, the evolution outcome changes dramatically.
Thus, in games such as Harmony, where cooperation is the only
equilibrium and the only rational outcome, rapid selection leads to
dominance of defectors. Similar non trivial phenomena arise in other
binary games and even in more complicated settings such as the
Ultimatum game. We conclude that the rate of selection is a key
element to understand and model the emergence of cooperation, and one
that has so far been overlooked.
\end{abstract}

\bigskip


\paragraph{Short Title} Selection Rate and Evolution of Cooperation

\newpage

\section{Introduction}

A fundamental, profound and broad-ranging unsolved question is how
cooperation among animals and humans has evolved
\citep{Darwin,Maynard,Pennisi}. From the point of view of natural
selection, the question to be answered is why cooperation may be a
better strategy for survival or reproduction than a more selfish
behavior. For this reason evolutionary game theory has been the
mathematical framework that has provided the deepest insights into
this issue \citep{Axelrod,Gintis,Nowak-Sigmund}. Simple games such as
the Prisoner's Dilemma \citep{Axelrod}, the Snowdrift game
\citep{Sugden} or the Stag-Hunt game \citep{Skyrms} have been the
subject of intense experimental and theoretical work along this line
\citep{Camerer}. One of the main achievements of this approach has
been to show that the emergence of cooperation is sensitive to
whether populations are well-mixed, such as in replicator dynamics
evolution \citep{Taylor,Hofbauer,Gintis}, or spatially structured
\citep*{zz,zz2,Page,Doebeli-Hauert}. Co-evolution of agents and
networks \citep{Maxi,Victor} and finite population effects
\citep{Nowak-Sasaki} are also relevant factors to take into account
inasmuch as they may enhance or hinder cooperation. However, none of
these approaches has considered the influence of different selection
rates so far, surely because ever since Darwin it has been
acknowledged that natural selection acts at a very slow pace.
Nevertheless, recent experiments show that this may not always be the
case \citep{Hendry-Kinnison,Hendry-Wenburg,Yoshida-Jones}: under
certain circumstances (e.g. strong predation or captive breeding)
evolution selects for a new trait in just a few generations.

Generally, research on evolutionary game theory is based on a
population of individuals or agents that interact by playing a game.
In the absence of spatial structure, it is posited that every agent
plays the game against every other one, and then reproduction
proceeds according to the payoffs earned during the game stage. For
large populations, this amounts to saying that every player gains the
payoff of the game averaged in the current distribution of
strategies. In terms of time scales, such an evolution corresponds to
a regime in which selection takes place at a much slower rate than
the interaction between agents. However, these two time scales need
not be different in general and, in fact, for many specific
applications they can arguably be of the same order
\citep{Hendry-Kinnison,Hendry-Wenburg,Yoshida-Jones}.

Our main aim in this work is to show that the pace at which selection
acts on the population is crucial for the appearance and stability of
cooperation. Even in non-dilemma games such as Harmony
\citep{harmony}, where cooperation is the only possible rational
outcome, defectors may be selected for if population renewal is very
rapid. Similar results hold true for several others games, thus
pointing out the necessity to include a discussion on the rate of
selection, compared to the rate of interaction, in any study about
cooperation or any other situation modelled by evolutionary games.

\section{Setting: Dynamics and games}

When selection acts at a much slower rate than interaction, a widely
used tool to analyze frequency-dependent selection without mutation
is replicator dynamics \citep{Taylor,Hofbauer,Gintis}. Replicator
dynamics assumes a well-mixed population where all agents interact
before selection, with a per capita growth rate of each strategy
proportional to its fitness (the payoff earned in a round of games
between selection events). However, as stated in the previous
section, we aim to understand the effects of different selection
rates. To this end, we introduce the following new dynamics: There is
a population with $N$ players. A pair of individuals is randomly
selected for playing, earning each one an amount of fitness according
to the rules of the payoff matrix of the game. This game act is
repeated $s$ times, choosing a new random pair of players in each
occasion.

After every $s$ games, selection takes place. Following
 \citet{Nowak-Sasaki}, we have chosen Moran dynamics \citep{Moran}
to model selection in a finite population. One individual among
the population of $N$ players is chosen for reproduction
proportionally to its fitness, and its offspring replaces a
randomly chosen individual. As the fitness of all players is set
to zero before the following round of $s$ games, the overall
result is that all players have been replaced by one descendant,
but the player selected for reproduction has had a reproductive
advantage of doubling its offspring at the expense of the randomly
selected player. It is worth noting that the population size $N$
is therefore constant along the evolution.

The parameter $s$ controls the time scales of the model, i.e.
reflects the relation between the rate of selection and the rate of
interaction. For $s\ll N$ selection is very fast and very few
individuals interact between selection events. Higher values of $s$
represent proportionally slower rates of selection. Thus, when $s\gg
N$ selection is very slow and population is effectively well-mixed.

It only remains to specify the games we will be studying in this
paper, namely binary games. These have been widely used in
evolutionary game theory, because of their simplicity and
amenability for both analytical treatment and computer simulation.
A symmetrical binary game is completely defined by its $2\times2$
payoff matrix (or normal form, see e.g.\ \citet{Gintis})
\begin{equation}
\label{binary-game-matrix}
\begin{array}{ccc} \ & {\rm C} & {\rm D} \\
\begin{array}{c} {\rm C} \\ {\rm D} \end{array} &
\left(\begin{array}{c} a \\ b \end{array}\right. & \left.
\begin{array}{c} c \\ d \end{array}\right), \end{array}
\end{equation}
whose rows give the payoff obtained by each strategy when confronted
to all others including itself. When interpreted as a model for the
emergence of cooperation, C and D denote, respectively, the
strategies "cooperate" and "defect". Different kinds of games have
been defined according to the relations between the four coefficients
of the matrix \citep{Rapoport}.

\section{Results}

The dynamics we introduced in the previous section can be studied
analytically and exactly in the following manner: Let us denote by
$0\leq n\leq N$ the number of cooperators present in the population.
We calculate the probability $x_n$ of ending up in state $n=N$ (i.e.,
all players cooperate) when starting in state $n<N$. For $s=1$ and
$s\to\infty$, we obtain an exact, analytical expression for $x_n$.
For arbitrary values of $s$, such a closed form cannot be found;
however, it is possible to carry out a combinatorial analysis of the
possible combinations of rounds and evaluate, numerically but
exactly, $x_n$. See appendix \ref{maths} for the detailed
mathematics.

As our first and most striking example of the influence of the
selection rate, we will start by considering the Harmony game
\citep{harmony}, determined by $a>c>b>d$. The {\em only} Nash
equilibrium of this game is $({\rm C},{\rm C})$ , as it is obvious
from the payoffs: The best option for both players is to cooperate,
which yields the maximum payoff for each one. When this game is
framed in our dynamical model, Fig.\ \ref{fig:1}a shows that the
rationally expected outcome, namely that the final population
consists entirely of cooperators, is not achieved for small and
moderate values of the selection rate parameter $s$. For the smallest
values, only when starting from a population largely formed by
cooperators there is some chance of reaching full cooperation; most
of the times, defectors will eventually prevail and invade the whole
population. This counterintuitive result may arise even for values of
$s$ comparable to the population size, by choosing suitable payoffs
(not shown). Interestingly, the main result that defection is
selected for small values of $s$ does not depend on the population
size $N$; only details such as the shape of the curves (cf.\ Fig.\
\ref{fig:1}b) are modified by $N$.

[Figure \ref{fig:1} here]

In the preceding paragraph we have chosen the Harmony game to discuss
the effect of the rate of selection, but this effect is very general
and appears in many other games. To see this, consider the example of
the Stag-Hunt game \citep{Skyrms}, with payoffs $a>c>d>b$. This is
the paradigmatic situation of a $2\times 2$ game with two Nash
equilibria in pure strategies, one Pareto-dominant \citep{Gintis}
$({\rm C},{\rm C})$, in which players maximize their payoffs, and the
other risk-dominant \citep{Gintis} $({\rm D},{\rm D})$, in which
players minimize the possible damage resulting from a defection of
the partner. Which of these equilibria is selected has been the
subject of a long argument in the past, and rationales for both of
them can be provided. As Fig.\ \ref{fig:2} shows, results for small
$s$ are completely different from those obtained for larger values.
Indeed, we see that for $s=1$, all agents become defectors except for
initial densities of cooperators close to 1. However, for values of
$s \gtrsim N$ the resulting curve is quasi-symmetrical, reflecting a
much more balanced competition between both strategies.

[Figure \ref{fig:2} here]

Yet another example of the importance of the selection rate is
provided by the Snowdrift game \citep{Sugden}, defined by the
payoffs $c>a>b>d$. Also known as Chicken or Hawk-Dove, it is a
dilemma game not unrelated to, but different from, the Prisoner's
dilemma. Fig.\ \ref{fig:3} shows that for small values of $s$
defectors are selected for almost any initial fraction of
cooperators. When $s$ increases, we observe an intermediate regime
where both full cooperation and full defection have a nonzero
probability, which, interestingly, is almost independent of the
initial population. And, for large enough $s$, full cooperation is
almost always achieved.

[Figure \ref{fig:3} here]

Finally, we show for completeness the results for the most ubiquitous
game in studies about the evolution of cooperation: the Prisoner's
Dilemma \citep{Axelrod}, with payoffs $c>a>d>b$. As is well known, in
this game the rational choice is to defect. The effect of a small
number of games, shown in Fig.\ \ref{fig:4}, is to bias the game even
more towards defection. 
In any case, the initial density of cooperators must be very large
for cooperation to have some chance of becoming the selected
strategy, but for small values of $s$ this requirement is most
severe. Therefore, the parameter $s$ does not change the qualitative
behavior of the Prisoner's Dilemma, alhtough once again low $s$ works
against cooperation.

[Figure \ref{fig:4} here]

Similar results are to be found in almost any binary game. Indeed, it
can be shown that D strategists are selected for when the payoffs
satisfy $b<c$ in the extreme case $s=1$, irrespective of $a$, $d$, or
the population size $N$ (cf.\ Eq.\ (\ref{3b}) in appendix
\ref{maths}). Larger $s$ values can not be analyzed in such a simple
manner and, in particular, the corresponding results depend on all
four parameters of the payoff matrix.

The reason for this dramatic influence of the parameter $s$ resides
in the resulting fitness distribution over the population. For large
values of $s$ $(s \gtrsim N)$, most agents have played
between selection events, 
and as a consequence most of the population has nonzero payoff, which
in turn implies a nonzero probability of being selected for
reproductive advantage. On the contrary, with small $s$ $(s \ll N)$
the distribution concentrates on the few players that have actually
played, these ones being the only candidates for selection. This
results in a completely different probabilistic scenario, with the
consequences reflected in the results above.

As a specific example, consider the Stag-Hunt game shown in Fig.\
\ref{fig:2}, for the particular initial number of cooperators
$n=N/2$. With values of $s \gtrsim 100$, the payoffs are distributed
to a large set of pairs of players, with approximate frequencies of
$1/4$ $({\rm C},{\rm C})$, $1/4$ $({\rm D},{\rm D})$ and $1/2$ $({\rm
C},{\rm D})$ or $({\rm D},{\rm C})$. Given that the payoff matrix
fulfills $a+b=c+d$, both strategies collect a practically equal
amount of fitness, thus with no reproductive advantage for any
strategy and then $x_n \approx 0.5$. However, considering the case of
$s \ll 100$, only a few pairs are selected to play. With pairs $({\rm
C},{\rm C})$ or $({\rm D},{\rm D})$, the reproductive advantage is
obviously the same, as only one strategy receives all the fitness.
But pairs $({\rm C},{\rm D})$ or $({\rm D},{\rm C})$ will draw a
strong advantage to defectors, given the relation $b/c = 1/5$, which
in the end causes $x_n \approx 0$. Results for other games can be
understood in a similar way.

\section{Discussion and conclusion}

Let us now summarize our main findings. We have shown that selection
rate plays a crucial role in determining the fate of cooperation, by
studying how evolutionary dynamics in a Moran setting depends on the
number of times the game is played between selection events. We have
seen that even in a game as simple as Harmony, where cooperating is
the only rational outcome, rapid selection leads to the success of
defectors. We have observed similar behavior in other examples such
as the Snowdrift and the Stag-Hunt games. In binary games about
cooperation situations (including Prisoner's Dilemma), we have found
that rapid selection rates generally lead to the promotion of
defectors. In other contexts the interpretation of the results would
be different (such as a choice of the risk-dominant coordination
option in the Stag-Hunt game) but the effect of the selection rate
will undoubtedly be there. It is important to stress that the results
are fully analytic, involve no approximation, and apply to both
finite and infinite populations.

Although in this paper we have worked only in the framework of
binary games, we believe that our main claim, namely that
different time scales for interaction and selection can modify the
outcome of evolution, is relevant to evolutionary games in
general. Consider, for instance, the Ultimatum game
\citep*{Guth,Fehr-Fischbacher}, which is one of the most
frequently used games in theoretical and experimental studies of
cooperation. This game is much more complex than the previous
binary games, as it asymmetrical, i.e. each individual of the pair
of interacting players has a role, and there is a large number of
strategies, not just two. In the Ultimatum game, under conditions
of anonymity, two players are shown a sum of money. One of the
players, the ``proposer'', is asked to offer an amount of this sum
to the other, the ``responder''. The proposer can make only one
offer, which the responder can accept or reject. If the offer is
accepted, the money is shared accordingly (the proposer receives
the rest of the money); if rejected, both players gain nothing.
Since the game is played only once (no repeated interactions) and
anonymously (no reputation gain), a self-interested responder will
accept any amount of money offered whereas a self-interested
proposer will offer the minimum possible amount which will be
accepted. This is precisely the outcome predicted by the
replicator dynamics in a well-mixed population: the final
population will consist only of fully rational agents which offer
the smallest possible amount and accept any amount
\citep{Page-Nowak02}. On the contrary, relaxing the assumption of
a well-mixed population in favor of a dynamics that allows a
faster selection rate changes this result absolutely. Proceeding
similarly to the binary games of the previous section, pairs of
players are randomly selected to interact, in series of $s$ rounds
between reproduction-selection events. Exhaustive simulations of
different versions of the game with this kind of dynamics have
shown that fair split becomes then the dominant strategy
\citep{Sanchez}. An analytical study of this problem that confirms
the simulation results is under way \citep*{Miguel}.

Finally, the most important implication of our results is that, in
studies about the emergence of cooperation, the rate of selection
is an extremely influential parameter and very often leads to
non-trivial, unexpected outcomes. Of course, the scope of this
result is not limited to cooperation among humans, as cooperative
phenomena have been reported for many other species including
bacteria \citep*{bacteria}. In fact, the research reported here
stresses the importance of selection rates for evolutionary game
theory, for all the situations it models (not only cooperation)
and for evolutionary theory in general
\citep{Hendry-Kinnison,Hendry-Wenburg,Yoshida-Jones}.

\newpage

\appendix

\section{Exact analytical results}
\label{maths}

In Moran dynamics, at each time step an individual is chosen for
reproduction proportional to its fitness, and one identical offspring
is produced that replaces another randomly chosen individual. In a
population of $N$ individuals where $n$ are C strategists
(cooperators) and $N-n$ are D strategists (defectors), we have a
Markov process with a tridiagonal transition matrix (a birth-death
process \citep{karlin}) given by
\begin{eqnarray}
\label{2a}
P_{n,n+1}&=&\frac{N-n}{N}\left\langle
\frac{W^{\rm C}_n}{W^{\rm C}_n+W^{\rm D}_n}\right\rangle, \\
\label{2b}
P_{n,n-1}&=&\frac{n}{N}\left\langle
\frac{W^{\rm D}_n}{W^{\rm C}_n+W^{\rm D}_n}\right\rangle,
\end{eqnarray}
and $P_{n,n}=1-P_{n,n+1}-P_{n,n-1}$, where $W^{\alpha}_n$ is the
fitness earned by cooperators ($\alpha={\rm C}$) or defectors
($\alpha={\rm D}$) after $s$ games, and $\langle\cdot \rangle$
denotes the average over realizations of the process. The
dependence of the calculation on the selection rate, i.e., on $s$,
enters only in these two quantities.

The solution to this birth-death process is obtained in a standard
manner \citep{karlin}. Let us denote by $x_n$ the probability of
ending up in state $n=N$ when starting off from state $n$. Then we
have
\begin{equation}
\label{5} x_n=P_{n,n+1}x_{n+1}+P_{n,n}x_n+P_{n,n-1}x_{n-1}
\end{equation}
with boundary conditions $x_0=0$, $x_N=1$. The solution to this
equation is given by
\begin{equation}
\label{6}
x_n=\frac{Q_n}{Q_N}, \qquad Q_n=1+\sum_{j=1}^{n-1}
\prod_{i=1}^j\frac{P_{i,i-1}}{P_{i,i+1}}\quad (n>1),
\qquad Q_1=1.
\end{equation}
When the ratio of the transition probabilities can be written as
\begin{equation}
\label{7} \frac{P_{n,n-1}}{P_{n,n+1}}=\frac{\alpha
n+\beta}{\alpha(n+1)+\gamma},\qquad \alpha\ne 0,
\end{equation}
equation (\ref{6}) has the closed form
\begin{equation}
\label{8}
Q_n=\frac{\gamma}{\gamma-\beta}\left[1-\binom{n+(\beta/\alpha)}{n}
\binom{n+(\gamma/\alpha)}{n}^{-1}\right],
\end{equation}
where the generalized binomial coefficient is defined as $\binom{x}{n}
=x(x-1)\cdots(x-n+1)/n!$.

For the extreme case $s=1$ there are two possible outcomes of the
variable $W^{\rm C}_n/(W^{\rm C}_n+W^{\rm D}_n)$, namely $2a/2a$,
with probability $n(n-1)/N(N-1)$ (that of pairing two Cs), and
$b/(b+c)$, with probability $2n(N-n)/N(N-1)$ (that of pairing a C
and a D). Thus, we have
\begin{equation}
\label{3a}
\left\langle\frac{W^{\rm C}_n}{W^{\rm C}_n+W^{\rm D}_n}\right\rangle
=\frac{n}{N(N-1)}\left[\frac{c-b}{c+b}n+\frac{2b}{c+b}N-1\right].
\end{equation}
Using a similar reasoning for $W^{\rm D}_n/(W^{\rm C}_n+W^{\rm D}_n)$
we end up with
\begin{equation}
\label{3b}
\frac{P_{n,n-1}}{P_{n,n+1}}=\frac{(c-b)n+(c+b)N-c-b}{(c-b)n+2bN-c-b},
\end{equation}
which has the form (\ref{7}). Notice that if $b<c$ then
$P_{n,n+1}<P_{n,n-1}$ regardless the values of $a$ and $d$.

In the opposite limit $s\to\infty$ every player plays with every
other an infinite number of times, so
\begin{equation}
\lim_{s\to\infty}\left\langle\frac{W^{\alpha}_n}{W^{\rm C}_n
+W^{\rm D}_n}\right\rangle=\frac{\overline{W^{\alpha}_n}}
{\overline{W^{\rm C}_n}+\overline{W^{\rm D}_n}}, \qquad
\alpha=\mbox{C, D},
\end{equation}
where $\overline{W^{\alpha}_n}$ denotes the payoff of $\alpha$
strategists when every player plays with every other once. Thus,
in this case
\begin{eqnarray}
\label{4b}
\frac{P_{n,n-1}}{P_{n,n+1}}=\frac{(c-d)n+d(N-1)}{(a-b)n+bN-a},
\end{eqnarray}
which has the form (\ref{7}) only when $a+d=c+b$.

For other values of $s$, $W^{\rm C}_n$ and $W^{\rm D}_n$
can be computed through a combinatorial enumeration of all possible
pairings, and the resulting exact expression (too cumbersome to be
included here) can then be evaluated numerically.

\newpage

\paragraph{Acknowledgments}
We acknowledge financial support from Ministerio de Educaci\'on y
Ciencia of Spain through grants BFM2003-07749-C05-01,
FIS2004-01001 and NAN2004-09087-C03-03 (AS), BFM2003-0180 (JAC),
the Thematic Network FIS2004-22008-E and the Action
FIS2004-22783-E. Support from the Comunidad de Madrid Programme
for Research Groups at Universidad Carlos III and from the
European Science Foundation through the COST Action ``Physics of
Risk'' is also acknowledged.

\newpage

\begin{figure}[h]
\begin{center}
\includegraphics[width=10cm,angle=0,clip=]{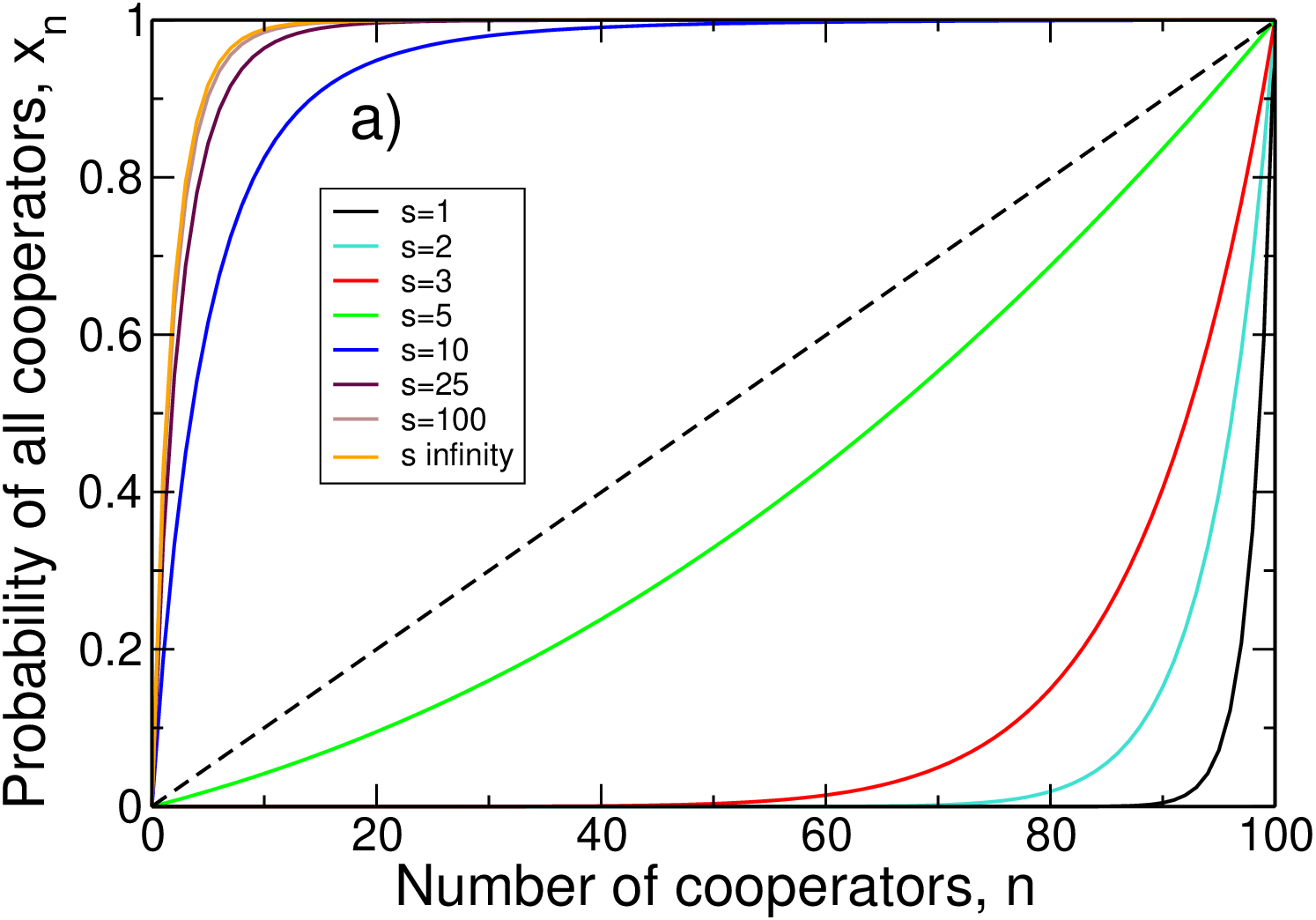}\\
\includegraphics[width=10cm,angle=0,clip=]{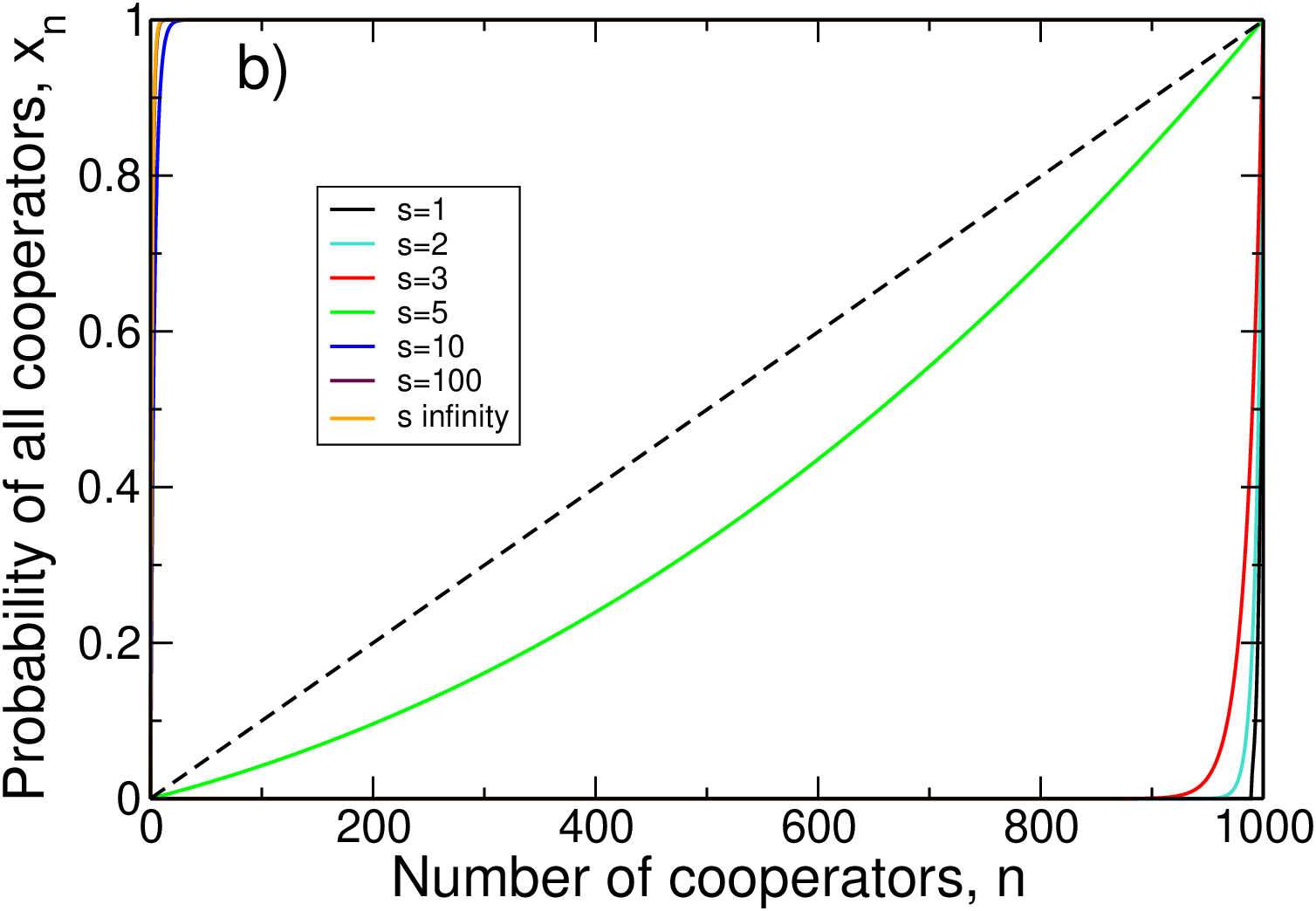}
\caption{\label{fig:1}}
\end{center}
\end{figure}

\paragraph{Figure \ref{fig:1}}
Probability $x_n$ of ending up with all cooperators starting from $n$
cooperators, for different values of $s$, in the Harmony game. a) For
the smallest values of $s$, full cooperation is only reached if
almost all agents are initially cooperators. Values of $s$ of the
order of 10 show a behavior much more favorable to cooperators. In
this plot, the population size is $N=100$. b) Taking a population of
$N=1000$, we observe that the range of values of $s$ for which
defectors are selected does not depend on the population size, only
the shape of the curves does. Parameter choices are: Number of games
between selection events, $s$, as indicated in the plots; payoffs for
the Harmony game, $a=11$, $b=2$, $c=10$, $d=1$. The dashed line
corresponds to a probability to reach full cooperation equal to the
initial fraction of cooperators and is shown for reference.

\newpage

\begin{figure}[h]
\begin{center}
\includegraphics[width=10cm]{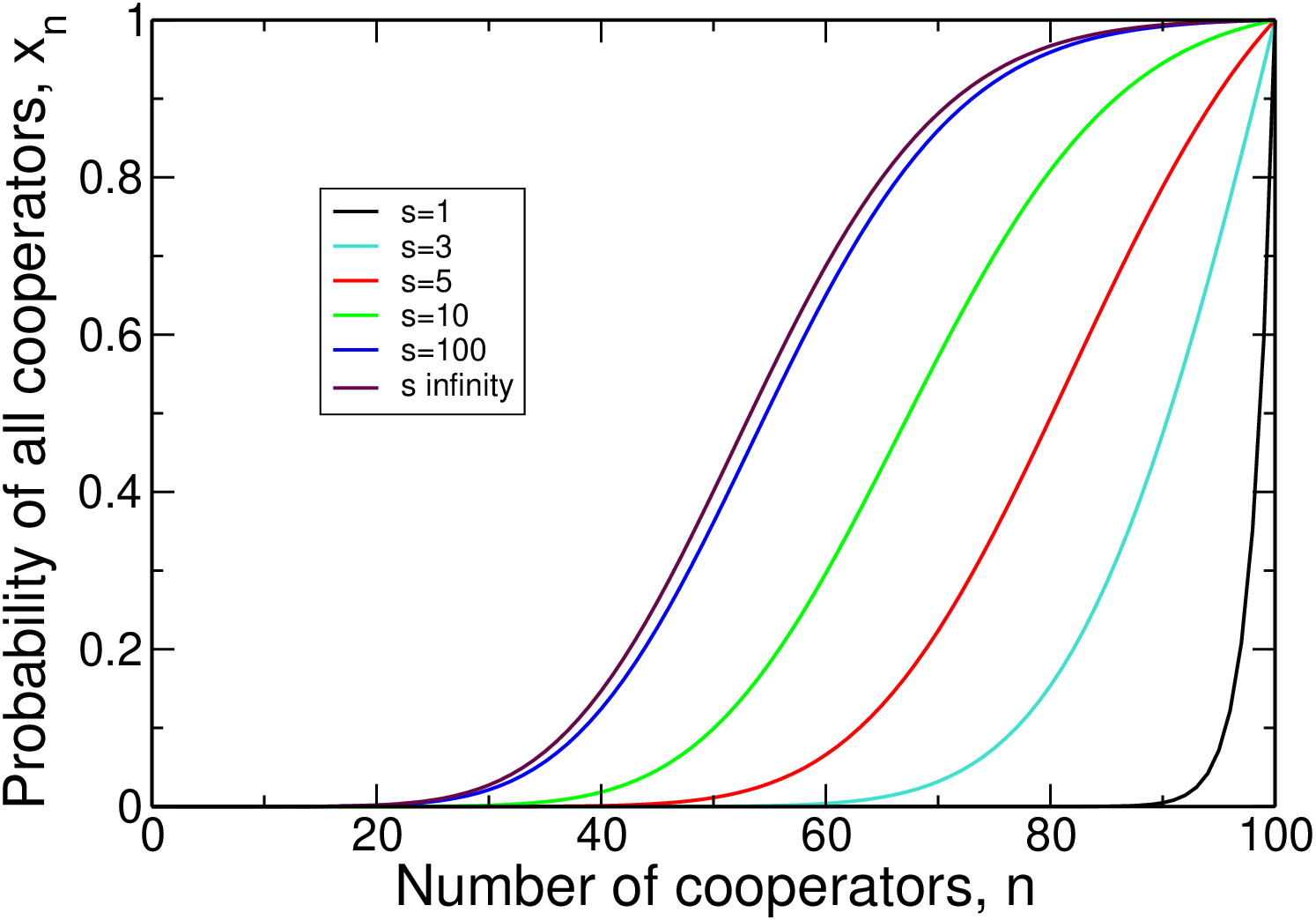}
\caption{\label{fig:2}}
\end{center}
\end{figure}

\paragraph{Figure \ref{fig:2}}
Same as Fig.\ \ref{fig:1}, for the Stag-Hunt game. The probability
$x_n$ of ending up with all cooperators when starting from $n$
cooperators, is very low if $s$ is small, and as $s$ increases it
tends to a quasi-symmetric distribution around $1/2$. Parameter
choices are: Population, $N=100$; number of games between selection
events, $s$, as indicated in the plot; payoffs for the Stag-Hunt
game, $a=6$, $b=1$, $c=5$, $d=2$.

\newpage

\begin{figure}[h]
\begin{center}
\includegraphics[width=10cm]{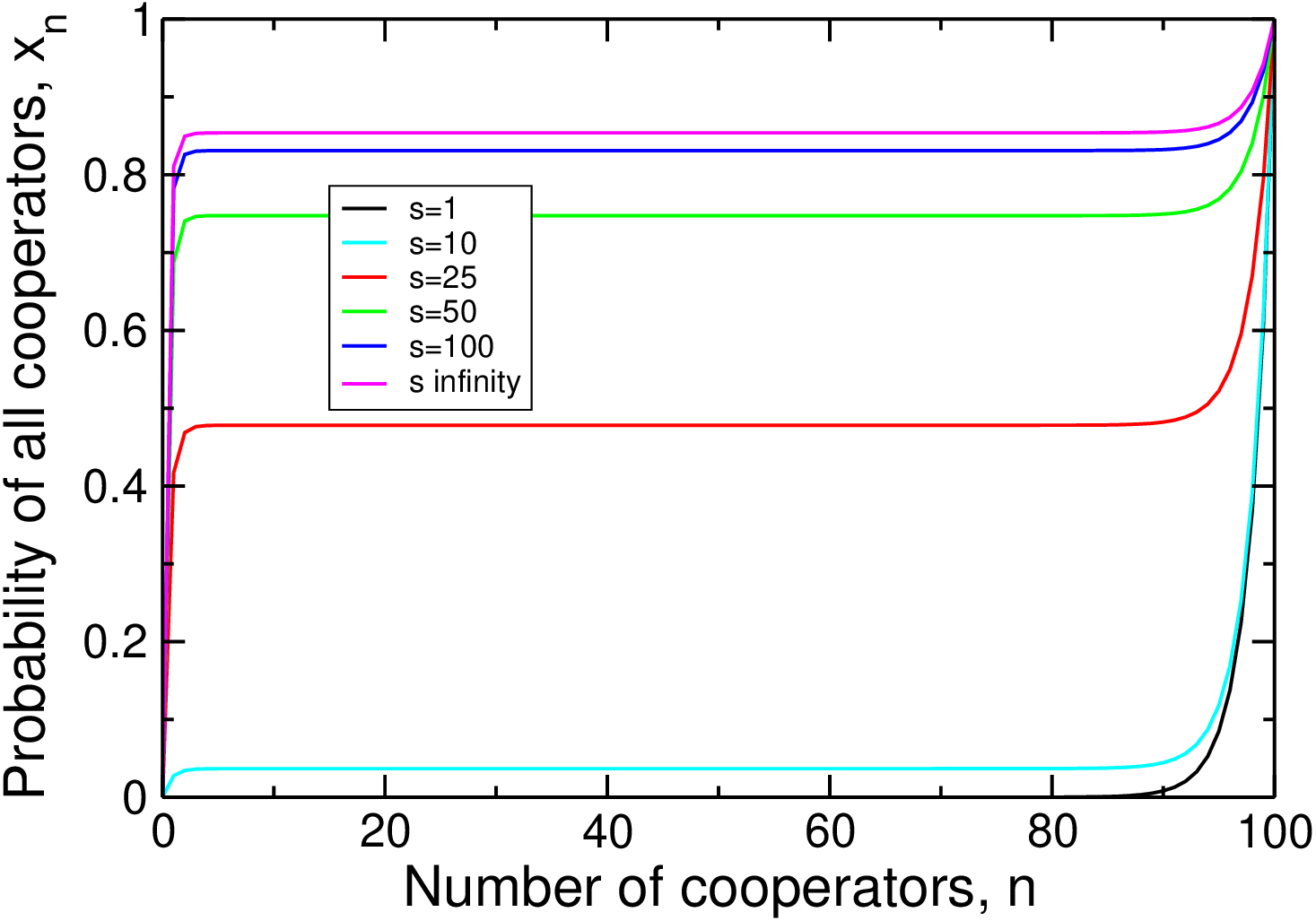}
\caption{\label{fig:3}}
\end{center}
\end{figure}

\paragraph{Figure \ref{fig:3}}
Same as Fig.\ \ref{fig:1}, for the Snowdrift game. The probability
$x_n$ of ending up with all cooperators starting from $n$ cooperators
is almost independent of $n$, except for very small or very large
values. Small $s$ values lead once again to selection of defectors,
whereas cooperators prevail more often as $s$ increases. Parameter
choices are: Population, $N=100$; number of games between selection
events, $s$, as indicated in the plot; payoffs for the Snowdrift
game, $a=1$, $b=0.35$, $c=1.65$, $d=0$.

\newpage

\begin{figure}[h]
\begin{center}
\includegraphics[width=10cm]{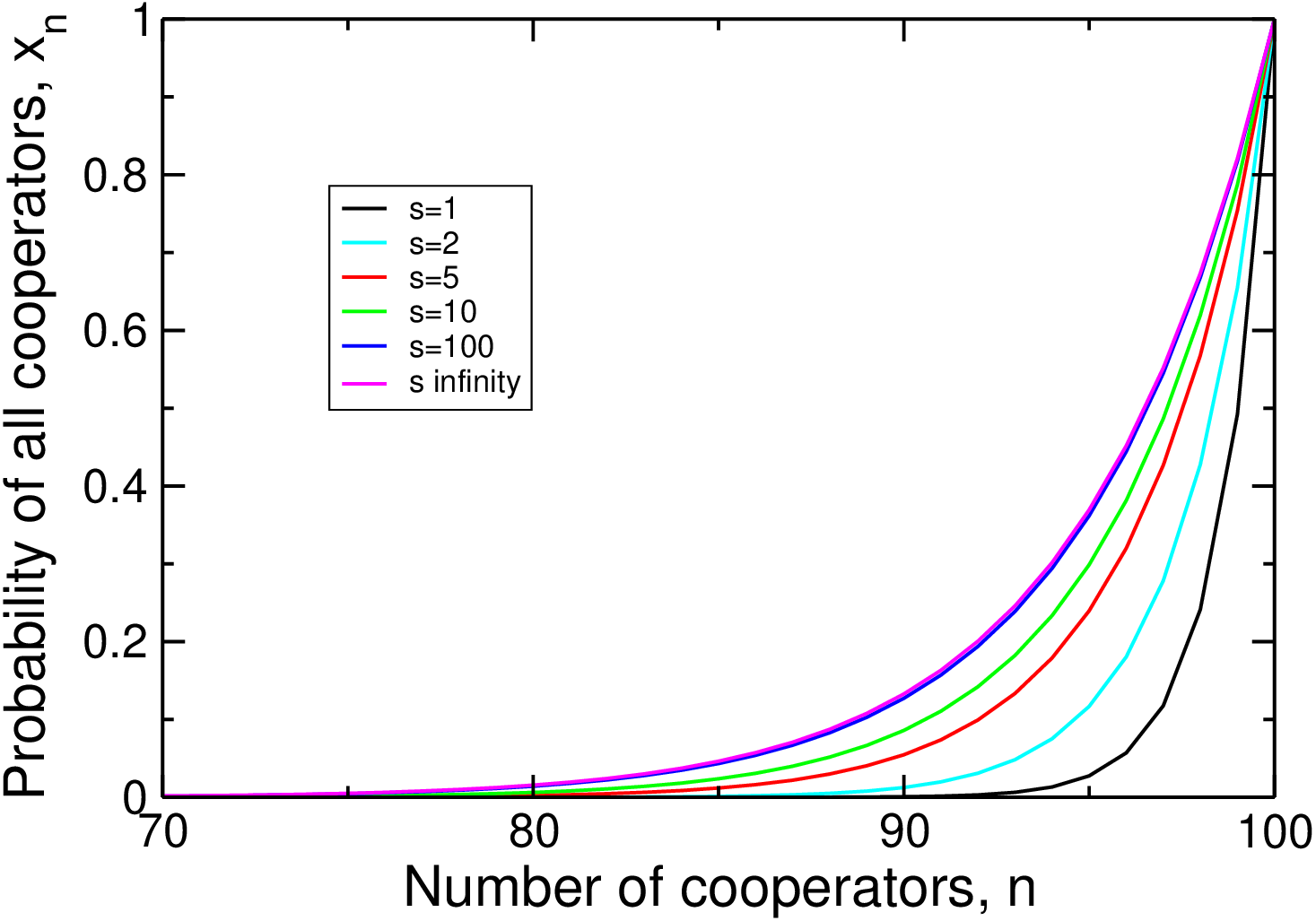}
\caption{\label{fig:4}}
\end{center}
\end{figure}

\paragraph{Figure \ref{fig:4}}
Same as Fig.\ \ref{fig:1}, for the Prisoner's Dilemma, but only
the rightmost part of the $n$ axis is shown (smallest $n$ values
have a negligible chance to give rise to cooperation). In this
game, small values of $s$ lead to an even larger possibility of
defection. Parameter choices are: Population, $N=100$; number of
games between selection events, $s$, as indicated in the plot;
payoffs for the Prisoner's Dilemma, $a=1$, $b=0$, $c=1.2$,
$d=0.1$.

\newpage

\newpage

\newpage

\newpage

\end{document}